\documentstyle[12pt]{article}

\def\beq{\begin{equation}}
\def\eeq{\end{equation}}
\def\bea{\begin{eqnarray}}
\def\eea{\end{eqnarray}}
\def\nn{\nonumber}
\def\sss{\scriptscriptstyle}

\def\bd{B_d^0}
\def\bdbar{{\overline{B_d^0}}}
\def\bs{B_s^0}
\def\bsbar{{\overline{B_s^0}}}
\def\barp{{\raise.35ex\hbox
{${\sss (}$}}---{\raise.35ex\hbox{${\sss )}$}}}
\def\bdbarp{\hbox{$B_d$\kern-1.4em\raise1.4ex\hbox{\barp}}}
\def\bsbarp{\hbox{$B_s$\kern-1.4em\raise1.4ex\hbox{\barp}}}
\def\ks{K_{\sss S}}
\def\kbar{{\overline{K^0}}}
\def\roughly#1{\mathrel{\raise.3ex\hbox
{$#1$\kern-.75em\lower1ex\hbox{$\sim$}}}}

\def\Abar{\bar A}
\def\Sbar{\bar S}
\def\thetaNP{\theta_{\sss NP}}

\def\ijmp#1#2#3{{\it Int.\ J.\ Mod.\ Phys.} {\bf A#1} (19#2) #3}

\def\plb#1#2#3{{\it Phys.\ Lett.} {\bf #1B} (19#2) #3}

\def\prd#1#2#3{{\it Phys.\ Rev.} {\bf D#1} (19#2) #3}

\def\prl#1#2#3{{\it Phys.\ Rev.\ Lett.} {\bf #1} (19#2) #3}

\def\zpc#1#2#3{{\it Zeit.\ Phys.} {\bf C#1} (19#2) #3}

\begin{document}
\setlength{\baselineskip}{20pt}

\begin{flushright}
UdeM-GPP-TH-99-61 \\
IMSc-99/05/18 \\
\end{flushright}

\begin{center}
\bigskip

{\Large \bf Can One Measure the Weak Phase}\\
{\Large \bf of a Penguin Diagram?}\\
\bigskip

David London $^{a,}$\footnote{london@lps.umontreal.ca},~~ 
Nita Sinha $^{b,}$\footnote{nita@imsc.ernet.in}~~
and Rahul Sinha $^{b,}$\footnote{sinha@imsc.ernet.in}
\end{center}


\begin{flushleft}
~~~~~~~~~~~$a$: {\it Laboratoire Ren\'e J.-A. L\'evesque, 
Universit\'e de Montr\'eal,}\\
~~~~~~~~~~~~~~~{\it C.P. 6128, succ. centre-ville, Montr\'eal, QC,
Canada H3C 3J7}\\
~~~~~~~~~~~$b$: {\it Institute of Mathematical Sciences, Taramani,
 Chennai 600113, India}
\end{flushleft}

\begin{center}
 
\bigskip (\today)

\bigskip 

{\bf Abstract}

\end{center}

\begin{quote}
  The $b\to d$ penguin amplitude receives contributions from internal
  $u$, $c$ and $t$-quarks. We show that it is impossible to measure
  the weak phase of any of these penguin contributions without
  theoretical input. However, it is possible to obtain the weak phase
  if one makes a single assumption involving the hadronic parameters.
  With such an assumption, one can test for the presence of new
  physics in the $b\to d$ flavour-changing neutral current by
  comparing the weak phase of $\bd$-$\bdbar$ mixing with that of the
  $t$-quark contribution to the $b\to d$ penguin.
\end{quote}
\newpage

\section{Introduction}

In the near future, it is expected that experiments at $B$-factories,
HERA-B, and hadron colliders will measure CP-violating rate
asymmetries in $B$ decays \cite{BCPasym}, thus yielding values of
$\alpha$, $\beta$ and $\gamma$, the three interior angles of the
unitarity triangle. What is particularly compelling about CP
violation in the $B$ system is that all three angles can be extracted
{\it cleanly}, i.e.\ without theoretical hadronic uncertainties. If
Nature is kind, these measurements will reveal the presence of physics
beyond the standard model (SM).

The most obvious way to detect new physics is to compare the unitarity
triangle as constructed from these CP angles with the triangle
constructed from independent measurements of the sides. Any
inconsistency will be evidence for new physics. The potential problem
with this approach is that there are large theoretical errors, all
related to hadronic physics, in extracting the lengths of the sides of
the unitarity triangle from the experimental data. Because of this,
the presently-allowed region for the unitarity triangle is still
rather large \cite{AliLondon}. Thus, it is conceivable that new
physics might be present, but we would still not be certain due to the
theoretical uncertainties.  Furthermore, even if the presence of new
physics were clearly established, this method would not tell us which
of the measurements of the sides and angles were affected by the new
physics.

In light of this, a more promising technique for searching for new
physics is to consider two distinct decay modes which, in the SM,
probe the same CP angle. If there is a discrepancy between the two
values, this would be unequivocal, clean evidence for new physics. In
addition, we would have a much better idea of where the new physics
entered.

In fact, there are several decay modes which can be used in this way.
For example, the angle $\gamma$ can be measured using rate asymmetries
in $B^\pm \to D K^\pm$ \cite{BDK} or $\bs(t) \to D_s^\pm K^\mp$
\cite{BsDsK}. And the angle $\beta$ can be measured via $\bd(t) \to
J/\Psi \ks$ or $\bd(t) \to \phi \ks$ \cite{LonSoni}. In either case, a
discrepancy in the values of the measured CP angles would be a
smoking-gun signal for new physics. (A third possibility, which is
similar in spirit to these two examples, is the CP asymmetry in
$\bs(t) \to J/\Psi \phi$. To a good approximation, in the SM this
asymmetry is zero, so that a nonzero value would clearly point to new
physics.)

If such a discrepancy were observed, what type of new physics could be
responsible? Tree-level weak decays, being dominated by $W$-exchange,
are essentially unaffected by new physics. Thus, new physics enters
principally through new contributions to loop-level processes, such as
$B^0$-${\overline{B^0}}$ mixing \cite{NPBmixing} or penguin decays
\cite{GroWorah}. We can therefore conclude that a discrepancy in the
value of $\gamma$ as extracted from $B^\pm \to D K^\pm$ and $\bs(t)
\to D_s^\pm K^\mp$ is due to the presence of new physics in
$\bs$-$\bsbar$ mixing. Similarly, since the decay $\bd(t) \to \phi \ks$
is a pure penguin process, new physics in the $b \to s$ penguin
amplitude can lead to different values of $\beta$ as measured in
$\bd(t) \to J/\Psi \ks$ and $\bd(t) \to \phi \ks$. (If there were new
physics in $\bd$-$\bdbar$ mixing, both of these decays would be
equally affected, so that this could not be the cause of any
discrepancy.) Thus, in both cases, not only would we be certain that
new physics is present, we would also know exactly where it had
entered. (Similarly, if the CP asymmetry in $\bs(t) \to J/\Psi \phi$
were found to be nonzero, this would clearly indicate the presence of
new physics in $\bs$-$\bsbar$ mixing.)

In all of these examples we are able to probe new physics in the $b
\to s$ flavor-changing neutral current (FCNC). The obvious question is
then: is there a way to use this type of method to probe new physics
in the $b \to d$ FCNC?

One possibility is to try to measure the weak phase of a $b \to d$
penguin diagram. In the (approximate) Wolfenstein parametrization
\cite{Wolfenstein} of the Cabibbo-Kobayashi-Maskawa (CKM) matrix, only
$V_{td}$ and $V_{ub}$ have significant non-zero phases. These phases
are two of the angles in the unitarity triangle: $\beta = {\rm
  Arg}(V_{td}^*)$ and $\gamma = {\rm Arg}(V_{ub}^*)$ ($\alpha$ is
defined to be $\pi - \beta - \gamma$). The $b\to d$ penguin amplitude
receives a contribution from an internal $t$-quark, and the product of
CKM matrix elements found in this contribution is $V_{tb}^* V_{td}$,
whose weak phase is $-\beta$. Thus, if one could compare the value of
$\beta$ as extracted from the $t$-quark contribution to the $b\to d$
penguin with that measured in some other decay (e.g.\ $\bd(t) \to
J/\Psi \ks$), one might be able to detect the presence of new physics
in the $b\to d$ FCNC.

If the $t$-quark contribution to the $b\to d$ penguin amplitude were
dominant, this would be straightforward. In this case, one could
simply measure CP violation in a pure $b\to d$ penguin decay such as
$\bd \to K^0 \kbar$ or $\bs \to \phi\ks$. In the SM, the CP asymmetry
in $\bd(t) \to K^0 \kbar$ is expected to vanish (the weak phase of
$\bd$-$\bdbar$ mixing cancels the weak phase of the $t$-quark penguin
amplitude), while the measurement of $\bs(t) \to \phi\ks$ allows one
to extract $\sin 2\beta$ \cite{LonPeccei}. If a disagreement were
found between these predictions and the experimental results, this
would be a clear indication of new physics in the $b\to d$ FCNC.

Unfortunately, things are not so easy. Theoretical estimates suggest
that the $b\to d$ penguin is {\it not} dominated by the internal
$t$-quark. On the contrary, the $u$- and $c$-quark contributions can
be substantial, perhaps even as large as 20\%--50\% of the $t$-quark
contribution \cite{ucquark}. If this is the case, the CP asymmetries
do not cleanly probe weak phases, and the SM predictions given above
are altered. The asymmetries now depend on (unknown) hadronic
quantities such as the strong phases and the relative sizes of the
various penguin contributions, so that a discrepancy between the above
predictions and the measurements does not necessarily imply new
physics.

Still, if we could find a way to {\it isolate} the $t$-quark
contribution to the $b\to d$ penguin, we could perhaps measure its
weak phase, and thereby test for the presence of new physics in the
$b\to d$ FCNC.

The main purpose of this paper is to examine whether such a method is
feasible. We will show that, in fact, it is impossible to {\it
  cleanly} measure the weak phase of the $t$-quark contribution to the
$b\to d$ penguin, or indeed the phase of any of the penguin
contributions. The reason is fundamentally very simple: due to the
unitarity of the CKM matrix, we have
\beq
V_{ub}^* V_{ud} + V_{cb}^* V_{cd} + V_{tb}^* V_{td} = 0 ~.
\label{unitarity}
\eeq
This is the equation used to define the unitarity triangle. But the
three terms in this equation are also the CKM matrix elements of the
$u$-, $c$- and $t$-quark contributions to the $b \to d$ penguin. It is
therefore impossible to isolate any one contribution -- it is always
possible to write a particular contribution in terms of the other two.
In this paper we refer to this as the ``CKM ambiguity.'' Since one
cannot isolate the $t$-quark contribution, it is clearly impossible to
measure its weak phase cleanly.

However, all is not lost. If the CKM ambiguity could somehow be
resolved, then it might be possible to measure the weak phase of the
$t$-quark contribution to the $b \to d$ penguin. In fact, as we will
show, this can be done, but it requires making a theoretical
assumption regarding the hadronic parameters of the penguin amplitude.
Since such an assumption holds only within a particular
parametrization of the penguin amplitude, this resolves the CKM
ambiguity, and allows us to extract the weak phase of the $t$-quark
$b\to d$ penguin, albeit not cleanly.

We discuss the CKM ambiguity in more detail in Sec.~2. We also show
explicitly that several methods which potentially could be used to
extract the weak phase of the $b \to d$ penguin in fact do not contain
enough information. In Sec.~3 we show that the CKM ambiguity can be
resolved by making a single assumption about the penguin parameters,
and give some examples of such assumptions. We conclude in Sec.~4.

\section{The CKM Ambiguity}

The full $b \to d$ penguin amplitude can be written as the sum of
three contributions:
\beq
P = \sum_{q=u,c,t} V_{qb}^* V_{qd} \, P_q ~,
\label{penguin}
\eeq
where we have explicitly separated out the dependence on the CKM
matrix element. In the Wolfenstein parametrization, the weak phases of
the $V_{ub}^* V_{ud}$, $V_{cb}^* V_{cd}$ and $V_{tb}^* V_{td}$ terms
are $\gamma$, 0 and $-\beta$, respectively.

Now, any one of the $V_{qb}^* V_{qd}$ terms can be eliminated using
Eq.~(\ref{unitarity}). Thus, depending on which term is eliminated,
there are several parametrizations one can use. For reasons which will
become clear below, we call this freedom the ``CKM ambiguity.''

Suppose, for example, that we choose to eliminate the $V_{ub}^*
V_{ud}$ piece. The penguin amplitude can then be written
\bea 
P & = & V_{cb}^* V_{cd} (P_c - P_u) + V_{tb}^* V_{td} (P_t - P_u) \nn \\ 
& \equiv & {\cal P}_{cu} \, e^{i\delta_{cu}} + {\cal P}_{tu} \,
e^{i\delta_{tu}} \, e^{-i\beta} ~,
\label{uelim}
\eea
where we have explicitly separated out the weak and strong phases and
absorbed the magnitudes $|V_{cb}^* V_{cd}|$ and $|V_{tb}^* V_{td}|$
into the definitions of ${\cal P}_{cu}$ and ${\cal P}_{tu}$,
respectively. We refer to this as parametrization \#1.

Suppose further that there exists a technique which permits us to
extract the weak phase $-\beta$ in the above expression {\it cleanly},
i.e.\ with no theoretical input regarding the remaining hadronic
parameters. If such a technique existed, then it would be possible to
express $-\beta$ (and perhaps the other parameters) entirely in terms
of measured observables.

However, if we had instead chosen to eliminate the $V_{tb}^* V_{td}$
piece from Eq.~(\ref{penguin}), we would have found
\bea 
P & = & V_{cb}^* V_{cd} (P_c - P_t) + V_{ub}^* V_{ud} (P_u - P_t) \nn \\ 
& \equiv & {\cal P}_{ct} \, e^{i\delta_{ct}} + {\cal P}_{ut} \,
e^{i\delta_{ut}} \, e^{i\gamma} ~,
\label{telim}
\eea
where ${\cal P}_{ct}$ and ${\cal P}_{ut}$ are defined in a similar
fashion to ${\cal P}_{cu}$ and ${\cal P}_{tu}$ in Eq.~(\ref{uelim})
above. We call this parametrization \#2.

The key point here is that parametrizations \#1 and \#2 are very
similar in form. If there existed a technique which could be used to
cleanly obtain $-\beta$ from parametrization \#1, that same technique
could be applied to parametrization \#2 to obtain $\gamma$.
Furthermore, the function of observables which gives $-\beta$ would be
the same function which yields $\gamma$, leading to the conclusion
that $-\beta = \gamma$, which is clearly false in general.

This argument demonstrates that it is impossible to cleanly measure
the weak phase of the $t$-quark contribution to the $b\to d$ penguin,
or indeed the weak phase of any contribution. This is due specifically
to the CKM ambiguity, i.e.\ the fact that the $b \to d$ penguin does
not have a well-defined parametrization. (A similar conclusion also
holds for the $b\to s$ penguin. However, in that case the situation is
slightly different. For the $b\to s$ penguin, the $c$- and $t$-quark
contributions are real in the Wolfenstein parametrization. And the
$u$-quark contribution, which has a nonzero weak phase, is
considerably suppressed relative to the others. Thus, to a good
approximation, one can say that the $b \to s$ penguin is real, so that
the CKM ambiguity is irrelevant.)

Even though this argument is quite conclusive, it is instructive to
examine several methods which one could conceivably use to attempt to
measure the weak phase of the $t$-quark contribution to the $b \to d$
penguin, and see exactly how they fail. In particular, we are
interested in counting the number of independent measurements, and
comparing this with the number of theoretical parameters. As we will
see, in all cases, the number of parameters exceeds the number of
measurements by one.

Before turning to specific examples, it is useful to establish some
notation. Due to $\bd$-$\bdbar$ mixing, a $\bd$ meson can evolve in
time into a mixture of $\bd$ and $\bdbar$. The time-dependent decay
rate for a $\bd(t)$ to decay into a final state $f$ is
\beq
\Gamma(\bd(t) \to f) = e^{-\Gamma t} \left[ {|A'|^2 + |\Abar'|^2 \over 2}
+ {|A'|^2 - |\Abar'|^2 \over 2} \cos (\Delta M t) 
- {\rm Im} \left( {q\over p} {A'}^* \Abar'\right) \sin (\Delta M t) \right],
\label{timedep}
\eeq
where $\bd(t)$ is a $B$-meson which at $t=0$ was a $\bd$, and $A'$ and
$\Abar'$ are $A(\bd \to f)$ and $A(\bdbar \to f)$, respectively. In
the Wolfenstein parametrization the mixing parameter $q/p$ takes the
form
\beq
{q\over p} = e^{- 2 i \beta}~.
\eeq

It is convenient to remove this mixing phase by redefining the
definitions of the decay amplitudes, i.e.\ 
\beq
A \equiv e^{i \beta} A' ~~,~~~~ \Abar \equiv e^{-i\beta} \Abar' ~.
\eeq
The time-dependent decay rate then allows us to extract $|A|$,
$|\Abar|$ and ${\rm Im}(A^* \Abar)$, i.e.\ the magnitudes of $A$ and
$\Abar$, as well as their relative phase.

In the examples which follow, we will adopt this notation, in which
the mixing phase has automatically been absorbed into the decay
amplitudes.

\subsection{$\bd(t) \to K^0 \kbar$}

The decay $\bd \to K^0 \kbar$ is a pure $b\to d$ penguin. The study of
the time-dependent decay rate for this decay allows one to obtain the
three quantities $|A|$, $|\Abar|$ and ${\rm Im}(A^* \Abar)$, where $A
\equiv e^{i\beta} A(\bd \to K^0 \kbar)$ and $\Abar \equiv e^{-i\beta}
A(\bdbar \to K^0 \kbar)$. However, it is straightforward to show that
this information alone does not allow us to extract any of the
theoretical parameters in the amplitudes.

Since the decay is pure penguin, the CKM ambiguity allows us to write
the amplitude $A$ in a variety of ways. Since we are interested in
measuring the weak phase of the $t$-quark contribution to the $b\to d$
penguin, we will keep the $V_{tb}^* V_{td}$ piece of the amplitude.
Suppose that we eliminate the $V_{ub}^* V_{ud}$ piece. The amplitude
$A$ can then be written
\bea 
A & = & e^{i \beta} \left[ {\cal P}_{cu} \, e^{i\delta_{cu}} 
  + {\cal P}_{tu} \, e^{i\delta_{tu}} \, e^{-i\beta'} \right] \nn\\
& = &   {\cal P}_{cu} \, e^{i\delta_{cu}} \, e^{i \beta} 
  + {\cal P}_{tu} \, e^{i\delta_{tu}} \, e^{-i\thetaNP} ~.
\label{BKKamplitude}
\eea
The quantities $\delta_{cu}$ and $\delta_{tu}$ are strong phases; only
their difference is measurable. Also, in the presence of new physics,
the phase of $\bd$-$\bdbar$ mixing may not be the same as that of the
$t$-quark contribution to the $b\to d$ penguin. We have allowed for
this possibility by writing the weak phase of the penguin as $\beta'$
in the first line. The new-physics phase is defined as $\thetaNP =
\beta' - \beta$. Measuring the phase of the $t$-quark penguin
contribution then is equivalent to measuring $\thetaNP$. The $\Abar$
amplitude can be obtained from the above equation simply by changing
the signs of the weak phases $\beta$ and $\thetaNP$.

From this expression we can count the number of theoretical
parameters. There are five: $\beta$, $\thetaNP$, ${\cal P}_{cu}$,
${\cal P}_{tu}$ and $\delta_{cu} - \delta_{tu}$. Since we have 3
measurements in 5 unknowns, it is impossible to solve for these
parameters. In particular, one cannot obtain $\thetaNP$. We can
improve things slightly by noting that $\beta$, which is the phase of
$\bd$-$\bdbar$ mixing, can be independently measured in $\bd(t) \to
J/\Psi \ks$. However, this still gives us one more unknown than there
are measurements.

If we had instead eliminated the $V_{cb}^* V_{cd}$ piece, this
conclusion would not change. Including now the independent measurement
of $\alpha$ (say in $\bd(t) \to \pi^+ \pi^-$), we would still be left
with 4 measurements in 5 unknowns.

In light of the CKM ambiguity it was to be expected that we would be
unable to cleanly extract $\thetaNP$. However, the point that we wish
to stress here is that there is only one more unknown than there are
measurements.

\subsection{Isospin Analysis of $B \to \pi\pi$}

The decay mode which is usually associated with the measurement of the
CP angle $\alpha$ is $\bd(t) \to \pi^+ \pi^-$. A decade ago it was
noticed that $b\to d$ penguin contributions, if large, can spoil the
clean extraction of $\alpha$ \cite{LonPeccei,penguins}. This is often
referred to as ``penguin pollution.'' Shortly thereafter, a method was
proposed for removing the penguin pollution. This method was based on
the fact that the amplitudes for the decays $\bd \to \pi^+ \pi^-$,
$\bd \to \pi^0 \pi^0$ and $B^+ \to \pi^+ \pi^0$ form a triangle in
isospin space \cite{isospin}.

In general, the decay $\bd \to \pi^+ \pi^-$ receives contributions
from a tree diagram and a $b\to d$ penguin diagram. Using unitarity to
eliminate the $V_{cb}^* V_{cd}$ piece of the penguin diagram, we can
write
\bea 
{1\over \sqrt{2}} \, A^{+-} & = & e^{i \beta} \left[ 
  - T^{+-} e^{i \delta^{+-}} e^{i \gamma} 
  + P e^{i \delta_{\sss P}} e^{-i\beta'} \right] \nn\\
& = & T^{+-} e^{i \delta^{+-}} e^{-i \alpha} 
  + P e^{i \delta_{\sss P}} e^{-i\thetaNP} ~.
\eea
In the above the $T^{+-} e^{i \delta^{+-}}$ term includes the
$u$-quark piece of the penguin amplitude, and $\delta^{+-}$ and
$\delta^{00}$ are strong phases.

The $\Abar^{+-}$ amplitude is obtained from the $A^{+-}$ amplitude by
changing the signs of the weak phases $\alpha$ and $\thetaNP$. If
there were no penguin contributions (i.e.\ $P=0$), then we would have
${\rm Im}({A^{+-}}^* \Abar^{+-}) \sim \sin 2\alpha$, so that we could
obtain a clean measurement of $\alpha$. However, if $P\ne 0$, then the
phase probed in ${\rm Im}({A^{+-}}^* \Abar^{+-})$ is clearly a
complicated function of $\alpha$ and the other parameters. Thus,
$\alpha$ can no longer be extracted cleanly.

The situation can be improved by using an isospin analysis. Isospin
relates the amplitude for $\bd \to \pi^+ \pi^-$ to the amplitudes for
$\bd \to \pi^0 \pi^0$ and $B^+ \to \pi^+ \pi^0$:
\beq
{1\over \sqrt{2}} \, A^{+-} + A^{00} = A^{+0} ~,
\label{Atriangle}
\eeq
with a similar triangle relation for the conjugate decays:
\beq
{1\over \sqrt{2}} \, \Abar^{+-} + \Abar^{00} = \Abar^{-0} ~.
\label{Abartriangle}
\eeq
The amplitudes $A^{00}$ and $A^{+0}$ can be explicitly written as
\bea
A^{00} & = & T^{00} e^{i \delta^{00}} e^{-i \alpha} 
  - P e^{i \delta_{\sss P}} e^{-i\thetaNP} ~, \nn\\
A^{+0} & = & \left[ T^{+-} e^{i \delta^{+-}} + 
   T^{00} e^{i \delta^{00}} \right] e^{-i \alpha} ~,
\eea
where again $\delta^{00}$ is a strong phase, and only the difference
of strong phases is measurable. The $\Abar$ amplitudes are again
obtained from the above expressions by changing the signs of the weak
phases.

The angle $\alpha$ can then found as follows. The magnitudes of the
six amplitudes $|A^{+-}|$, $|A^{00}|$, $|A^{+0}|$, $|\Abar^{+-}|$,
$|\Abar^{00}|$ and $|\Abar^{-0}|$, can be measured experimentally. We
can therefore construct the $A$- and $\Abar$-triangles
[Eqs.~(\ref{Atriangle}) and (\ref{Abartriangle})]. In addition, ${\rm
  Im}({A^{+-}}^* \Abar^{+-})$ gives the relative phase between the
$A^{+-}$ and $\Abar^{+-}$ amplitudes, thereby fixing the relative
orientations of the $A$- and $\Abar$-triangles. The key point is that
this then fixes the relative orientations of the $A^{+0}$ and
$\Abar^{-0}$ amplitudes. But the relative phase of these two
amplitudes is just $2\alpha$. Thus, the isospin analysis allows one to
remove the penguin pollution and cleanly extract $\alpha$. (In fact,
there are discrete ambiguities in the above procedure, but they are
not our concern here.)

Although it is nice to be able to obtain $\alpha$ cleanly, the
question which we wish to explore in this paper is: can we get more?
In particular, is there enough information to also extract $\thetaNP$?
It is straightforward to show that the answer is no. 

First, we note that there are a total of 7 parameters which appear in
the theoretical expressions for the amplitudes: $\alpha$, $\thetaNP$,
$T^{+-}$, $T^{00}$, $P$, $\Delta^{+-} \equiv \delta^{+-} -
\delta^{\sss P}$ and $\Delta^{00} \equiv \delta^{00} - \delta_{\sss
  P}$. Experimentally, at best one can measure the magnitudes and
relative phases of the six $A$ and $\Abar$ amplitudes, giving 11
measurements.  However, due to the $A$ and $\Abar$ triangle relations,
the four measurements involving the $A^{00}$ and $\Abar^{00}$
amplitudes are not independent. Furthermore, we have $|A^{+0}| =
|\Abar^{-0}|$. Thus, of the 11 measurements, only 6 are independent.
With 6 measurements in 7 unknowns, one cannot solve for $\thetaNP$. 

Again, given the discussion of the CKM ambiguity, this was to be
expected. However, as before, we find that there is only one more
unknown than there are measurements.

\subsection{Dalitz Plot Analysis of $B \to 3\pi$}

An alternative way to cleanly extract $\alpha$ in the presence of
penguin contributions is to study the Dalitz plot of $\bd(t)\to
\pi^+\pi^-\pi^0$ decays \cite{Dalitz}. This final state can be reached
via the intermediate states $\rho^+ \pi^-$, $\rho^-\pi^+$ and $\rho^0
\pi^0$. It is the interference between these intermediate states which
allows one to remove the penguin pollution and cleanly obtain
$\alpha$.

In this method, it is the $B \to \rho \pi$ amplitudes which are the
key ingredients. Isospin allows one to relate neutral $B\to \rho\pi$
decays to charged $B\to \rho\pi$ decays. Defining
\bea
S_1 & \equiv & e^{i\beta} \, \sqrt{2} A(B^+ \to \rho^+ \pi^0) ~, \nn\\
S_2 & \equiv & e^{i\beta} \, \sqrt{2} A(B^+ \to \rho^0 \pi^+) ~, \nn\\
S_3 & \equiv & e^{i\beta} \, A(\bd \to \rho^+ \pi^-) ~, \nn\\
S_4 & \equiv & e^{i\beta} \, A(\bd \to \rho^- \pi^+) ~, \nn\\
S_5 & \equiv & e^{i\beta} \, 2 A(\bd \to \rho^0 \pi^0) ~,
\eea
one can form an isospin pentagon:
\beq
S_1 + S_2 = S_3 + S_4 + S_5 ~.
\eeq
As in the $B\to \pi\pi$ case, there are in general both tree and $b\to
d$ penguin contributions to $B\to \rho\pi$ decays. Eliminating again
the $V_{cb}^* V_{cd}$ piece, the above amplitudes can be written
explicitly as follows \cite{Dalitz}:
\bea
S_1 & = & T^{+0} e^{i\delta^{+0}} e^{-i\alpha} 
             + 2 P_1 e^{i\delta_1} e^{-i \thetaNP} ~, \nn\\
S_2 & = & T^{0+} e^{i\delta^{0+}} e^{-i\alpha} 
             - 2 P_1 e^{i\delta_1} e^{-i \thetaNP} ~, \nn\\
S_3 & = & T^{+-} e^{i\delta^{+-}} e^{-i\alpha} 
             + P_1 e^{i\delta_1} e^{-i \thetaNP} 
             + P_0 e^{i\delta_0} e^{-i \thetaNP} ~, \nn\\
S_4 & = & T^{-+} e^{i\delta^{-+}} e^{-i\alpha} 
             - P_1 e^{i\delta_1} e^{-i \thetaNP} 
             + P_0 e^{i\delta_0} e^{-i \thetaNP} ~, \nn\\
S_5 & = & - T^{+-} e^{i\delta^{+-}} e^{-i\alpha} 
          - T^{-+} e^{i\delta^{-+}} e^{-i\alpha} 
          + T^{+0} e^{i\delta^{+0}} e^{-i\alpha} 
          + T^{0+} e^{i\delta^{0+}} e^{-i\alpha} 
          - 2 P_0 e^{i\delta_0} e^{-i \thetaNP} ~.
\eea     

There is a similar pentagon relation for the conjugate amplitudes:
\beq
\Sbar_1 + \Sbar_2 = \Sbar_3 + \Sbar_4 + \Sbar_5 ~,
\eeq
in which the $\Sbar_i$ amplitudes can again be obtained from the $S$
amplitudes by changing the signs of the weak phases $\alpha$ and
$\thetaNP$.

The Dalitz plot of the $\pi^+ \pi^- \pi^0$ final state contains enough
information to determine the magnitudes and relative phases of the six
amplitudes $S_3$, $S_4$, $S_5$, $\Sbar_3$, $\Sbar_4$ and $\Sbar_5$.
One can then obtain $\alpha$ via 
\beq
{S_3 + S_4 + S_5 \over \Sbar_3 + \Sbar_4 + \Sbar_5} = e^{-2 i \alpha} ~.
\eeq

As in the $B\to\pi\pi$ case, one can again show that there is not
enough information to extract $\thetaNP$. There are a total of 13
theoretical parameters: $\alpha$, $\thetaNP$, 6 $T$ and $P$
amplitudes, and 5 relative strong phases. Experimentally, one can
determine the magnitudes and relative phases of all $S$ and $\Sbar$
amplitudes ($S_1$, $S_2$, $\Sbar_1$ and $\Sbar_2$ can be obtained from
an analysis of the Dalitz plot of $\pi^+\pi^0\pi^0$). Thus, there are
nominally 19 measurements. However, 
\begin{itemize}
  
\item Due to the $S$ and $\Sbar$ pentagon relations, the amplitudes
  $S_5$ and $\Sbar_5$ are not independent. This removes 4
  measurements.
  
\item We have the equality $|S_1 + S_2| = |\Sbar_1 + \Sbar_2|$. This
  removes one more measurement.

\item It is easy to verify the complex equality
\beq
{S_3 - S_4 - S_1 \over \Sbar_3 - \Sbar_4 - \Sbar_1} = 
{S_1 + S_2 \over \Sbar_1 + \Sbar_2 } ~.
\eeq
This removes 2 more measurements.

\end{itemize}

Thus, of the 19 measurements, in fact only 12 are independent. Since
there are 13 unknowns, we cannot solve for $\thetaNP$, as per the CKM
ambiguity\footnote{We note that this contradicts one of the
  conclusions of Ref.~\cite{Dalitz}. The authors of Ref.~\cite{Dalitz}
  concede that this particular point is in error in their paper. We
  thank Helen Quinn for discussions of this matter.}. And, as in the
$B\to \pi\pi$ case there is one more unknown than there are
measurements.

\subsection{Angular Analysis of $B \to V V$ Decays}

Consider the case where a neutral $B$ meson decays to a final state
consisting of two vector mesons $V$. Due to the fact that this final
state does not have a well-defined orbital angular momentum, it cannot
be a CP eigenstate. However, it is possible to disentangle the CP-even
and CP-odd components of the $VV$ state through a helicity analysis of
the decay products of the $V$ mesons \cite{helicity}.

The amplitudes for a neutral $B$ or ${\bar B}$ meson to decay into a
pair of vector mesons can be written as
\bea
A(\displaystyle B\to V_1 V_2) & = & \sum_{\lambda =\left\{0,\perp,\|
\right\}} {\cal A}_\lambda  \zeta_\lambda f_\lambda ~, \nn\\
\Abar(\displaystyle {\bar B}\to V_1 V_2) & = & \eta_{\sss CP}
\sum_{\lambda = \left\{0,\perp,\| \right\}} 
\bar{\cal A} _\lambda \zeta^*_\lambda f_\lambda ~,
\eea
where the $f_\lambda$ are the coefficients of the helicity amplitudes
written in the linear polarization basis; the $f_\lambda$ depend only
on the angles describing the kinematics. The factor $\zeta_\lambda$
has been introduced to account for the fact that the $P$-wave
amplitude $A_\perp$ is $CP$-odd, while $A_0$ and $A_\|$ are $CP$-even.
Thus, $\zeta_\lambda=1$ for $\lambda = \left\{ 0,\| \right\}$ and
$\zeta_\lambda=i$ for $\lambda = \perp$. The intrinsic $CP$ parity of
the $A_0$ final state is defined as $\eta_{\sss CP}$.

The quantities which appear in the time-dependent decay rate
[Eq.~(\ref{timedep})] are
\beq
|A|^2=
\sum_{\lambda,\sigma} {\cal A}_\lambda {\cal A}_\sigma^* \zeta_\lambda 
\zeta_\sigma^* f_\lambda f_\sigma ~,
\label{Asq}
\eeq
\beq
|\Abar|^2=
\sum_{\lambda,\sigma} 
{\bar{\cal A}}_\lambda {\bar{\cal A}}_\sigma^* {\zeta^*_\lambda}
{\zeta_\sigma} f_\lambda f_\sigma ~,
\eeq
and
\beq
A^*{\bar A}= \sum_{\lambda,\sigma} 
{\cal A}_\lambda^* {\bar{\cal A}}_\sigma \zeta_\lambda^*
\zeta_\sigma^*  f_\lambda f_\sigma ~.
\label{Aint}
\eeq
Using Eqs.~(\ref{Asq})-(\ref{Aint}), Eq.~(\ref{timedep}) can be
rewritten as
\bea
\Gamma(\bd(t)\to f)=\displaystyle e^{-\Gamma t}\sum_{\lambda\leq\sigma}
\Bigl(\Lambda_{\lambda\sigma}+\Sigma_{\lambda\sigma}\cos(\Delta M t)
-\rho_{\lambda\sigma}\sin(\Delta M t)\Bigr) f_\lambda f_\sigma ~,
\label{Amp}
\eea
where the summation is done realizing the fact that $f_\lambda
f_\sigma$ cannot be distinguished from $f_\sigma f_\lambda$ in an
angular analysis. The quantities appearing in the above equation are
defined as
\bea
\Lambda_{\lambda\lambda} & = & \frac{|{\cal A}_\lambda|^2+|{\bar{\cal 
A}}_\lambda|^2}{2} ~,
\label{Lambda11}\\
\Lambda_{\lambda\sigma} & = & {\rm Re}({\cal A}_\lambda {\cal A}_\sigma^* 
\zeta_\lambda \zeta_\sigma^*+ {\bar{\cal A}}_\lambda {\bar{\cal 
A}}_\sigma^* \zeta_\lambda^* \zeta_\sigma ) ~, ~~~~~~~~~~\lambda\neq\sigma ~,
\label{Lambda}\\
\Sigma_{\lambda\lambda} & = & \frac{|{\cal A}_\lambda|^2-|{\bar{\cal 
A}}_\lambda|^2}{2} ~, \label{Sigma11}\\
\Sigma_{\lambda\sigma} & = & {\rm Re}({\cal A}_\lambda {\cal A}_\sigma^* 
\zeta_\lambda \zeta_\sigma^*- {\bar{\cal A}}_\lambda {\bar{\cal 
A}}_\sigma^* \zeta_\lambda^* \zeta_\sigma ) ~, ~~~~~~~~~~\lambda\neq\sigma ~,
\label{Sigma}\\
\rho_{\lambda\lambda} & = & {\rm Im}\Bigl({\cal A}_\lambda^* 
{\bar{\cal A}}_\lambda\,\zeta_\lambda^{*2}\Bigr) ~,
\label{rho11} \\               
\rho_{\lambda\sigma} & = & {\rm Im}\Bigl(({\cal A}_\lambda^* 
{\bar{\cal A}}_\sigma+{\cal A}_\sigma^* {\bar{\cal 
A}}_\lambda)\,\zeta_\lambda^*\zeta_\sigma^*\Bigr) ~, ~~~~~~~~~~\lambda\neq\sigma ~.
\label{rho}
\eea
We remind the reader that we have adopted a notation in which the
mixing phase $q/p$ has been absorbed into the decay amplitudes ${\cal
  A}_\lambda$ and ${\bar{\cal A}}_\lambda$.

It is clear from Eqs.~(\ref{Amp})-(\ref{rho}) that 18 quantities can
be measured. However, it is equally clear that only 11 of these
observables are independent. The fundamental quantities are the six
amplitudes ${\cal A}_\lambda$ and ${\bar{\cal A}}_\lambda$, $\lambda =
0, \perp, \|$. The most one can measure is their magnitudes and
relative phases, for a total of 11 independent measurements. And in
fact, it is straightforward to show that the observables in
Eqs.~(\ref{Lambda11})-(\ref{rho}) suffice to measure these 11
quantities, up to a two-fold discrete ambiguity in the relative
phases.

What can we learn from this information? For definitiveness, let us
consider the pure $b\to d$ penguin decay $\bd \to K^* {\bar K}^*$,
which is quite similar to the previous decay $\bd \to K^0 \kbar$.
Using CKM unitarity to eliminate the $V_{ub}^* V_{ud}$ piece, the
helicity amplitudes can be written
\beq
{\cal A}_\lambda = 
  {\cal P}_{cu}^\lambda e^{i\delta_{cu}^\lambda} e^{i \beta} 
  + {\cal P}_{tu}^\lambda e^{i\delta_{tu}^\lambda} e^{-i\thetaNP} ~.
\label{uelimhelicity}
\eeq
As usual, the ${\bar{\cal A}}_\lambda$ amplitudes are obtained by
changing the signs of $\beta$ and $\thetaNP$ in the above expression.

We can now count the number of theoretical parameters in the decay
amplitudes. There are 13: $\beta$, $\thetaNP$, 6 ${\cal P}$
magnitudes, and 5 relative strong phases. Assuming that $\beta$ is
independently measured, this still leaves 12 measurements in 13
unknowns. Once again, we cannot obtain $\thetaNP$ cleanly. And once
again, there is one more theoretical unknown than there are
measurements.

\subsection{Isospin $+$ Angular Analysis of $B \to \rho\rho$ Decays}

As a final example, one can imagine combining isospin and angular
analyses. Consider the decay $\bd\to\rho^+\rho^-$. The $\rho^+ \rho^-$
final state is not a CP eigenstate. An angular analysis can
distinguish the CP-even piece from the CP-odd piece by separating out
the three helicities. That is, we could obtain the magnitudes and
relative phases of the amplitudes $A^{+-}_\lambda$ ($\lambda = 0,
\perp, \|$), along with the corresponding conjugate amplitudes
$\Abar^{+-}_\lambda$, where
\beq 
{1\over \sqrt{2}} \, A^{+-}_\lambda = T^{+-}_\lambda e^{i \delta^{+-}_\lambda} 
e^{-i \alpha} + P_\lambda e^{i \delta_{\sss P}^\lambda} e^{-i\thetaNP} ~.
\eeq
However, as in the $\bd \to \pi^+\pi^-$ case, in the presence of
penguin contributions, this is not enough to obtain $\alpha$ cleanly
-- an isospin analysis is also necessary.

Imagine, then, that an angular analysis were also performed on the
decays $\bd(t)\to\rho^0\rho^0$ and $B^+ \to \rho^+ \rho^0$. We could
then also obtain the amplitudes $A^{00}_\lambda$, $A^{+0}_\lambda$,
along with their conjugate amplitudes. These amplitudes can be written
as
\bea
A^{00}_\lambda & = & T^{00}_\lambda e^{i \delta^{00}_\lambda} e^{-i \alpha} 
  - P_\lambda e^{i \delta_{\sss P}^\lambda} e^{-i\thetaNP} ~, \nn\\
A^{+0}_\lambda & = & \left[ T^{+-}_\lambda e^{i \delta^{+-}_\lambda} + 
   T^{00}_\lambda e^{i \delta^{00}_\lambda} \right] e^{-i \alpha} ~.
\eea
The $\Abar_\lambda$ amplitudes are obtained by changing the signs of
the weak phases. The helicity amplitudes form isospin triangles:
\bea
{1\over \sqrt{2}} \, A^{+-}_\lambda + A^{00}_\lambda & = & A^{+0}_\lambda ~,
{1\over \sqrt{2}} \, \Abar^{+-}_\lambda + \Abar^{00}_\lambda = \Abar^{-0}_\lambda ~.
\eea
There are thus 6 isospin triangles involving 18 amplitudes.

From the above, we see that there are a total of 19 theoretical
parameters: $\alpha$, $\thetaNP$, 9 magnitudes ($T_\lambda^{+-}$,
$T_\lambda^{00}$, $P_\lambda$), and 8 strong-phase differences.
Experimentally, the magnitudes and relative phases of all 18
amplitudes can be obtained, giving a total of 35 measurements.
However, not all measurements are independent:
\begin{itemize}
  
\item Due to the isospin triangles, the amplitudes $A^{00}_\lambda$
  and $\Abar^{00}_\lambda$ are not independent. This removes 12
  measurements.
  
\item We have
\beq
{A_\lambda^{+0} \over \Abar_\lambda^{-0}} = e^{-2 i \alpha} 
~~,~~~~\lambda=0, \perp, \| ~.
\eeq
Thus, the magnitudes of $A_\lambda^{+0}$ and $\Abar_\lambda^{-0}$ are
equal, as are their relative phases, for the helicities $\lambda = 0,
\perp, \|$.  This removes an additional 5 measurements.

\end{itemize}

We therefore find that we have a total of 18 independent experimental
measurements, but 19 theoretical unknowns. Although we can cleanly
find the CP-phase $\alpha$, we cannot determine any of the remaining
parameters, including $\thetaNP$.

\section{Resolving the CKM Ambiguity}

In the previous section, we showed that, due to the CKM ambiguity, it
is not possible to cleanly measure the weak phase of a penguin
amplitude. Indeed, in all the examples considered, we found that there
were more theoretical parameters than measurements, in agreement with
this result. However, what is interesting about the study of these
examples is that in all cases there was only one more unknown than
there were measurements. Although we did not present a proof, the
result appears to be very general. 

This result indicates something quite useful: if we wish to test for
the presence of new physics in the $b\to d$ FCNC by comparing the weak
phase of $\bd$-$\bdbar$ mixing with that of the $t$-quark contribution
to the $b\to d$ penguin, it is necessary to make a single assumption
about the theoretical (hadronic) parameters describing the decay. This
assumption will hold in only one parametrization of the decay
amplitude, and will therefore resolve the CKM ambiguity. Furthermore,
the requirement of a single theoretical assumption holds regardless of
which type of method is used.

In this section, we present several examples which show explicitly how
a theoretical assumption allows one to extract the weak phase of the
$t$-quark contribution to the $b\to d$ penguin, thereby enabling one
to test for the presence of new physics in the $b\to d$ FCNC.

\subsection{$\bd(t) \to K^0 \kbar$ and $\bs(t) \to \phi\ks$}

As mentioned in the introduction, $\bd \to K^0 \kbar$ and $\bs \to
\phi\ks$ are pure $b\to d$ penguin decays. It was recently shown in
Ref.~\cite{KLY} that, by measuring the time-dependent decay rates for
these processes, and adding a theoretical assumption, together these
decays can be used to measure the weak phase of the $t$-quark
contribution to the $b\to d$ penguin.

The amplitude for $\bd \to K^0 \kbar$ was given in
Eq.~(\ref{BKKamplitude}) and is repeated for convenience below:
\bea 
A_d^{\sss KK} = {\cal P}_{cu} \, e^{i\delta_{cu}} \, e^{i \beta} 
  + {\cal P}_{tu} \, e^{i\delta_{tu}} \, e^{-i(\beta'-\beta)} ~,
\label{BdKK}
\eea
where $\beta$ is the weak phase of $\bd$-$\bdbar$ mixing and $\beta'$
is the weak phase of the $t$-quark contribution to the $b\to d$
penguin.

The amplitude for $\bs \to \phi\ks$ can be written
\beq
A_s^{\sss \phi K} = \tilde{\cal P}_{cu} \, e^{i\tilde\delta_{cu}}
+ \tilde{\cal P}_{tu} \, e^{i\tilde\delta_{tu}} \, e^{-i\beta'} ~,
\label{BsphiK}
\eeq
where we have assumed that there is no new physics in $\bs$-$\bsbar$
mixing. (As discussed previously, it is possible to directly test for
the presence of such new physics. If it turns out that there is new
physics in $\bs$-$\bsbar$ mixing, it can be included straightforwardly
in the above equation.)

Comparing the above two equations, one immediately notes that the
parameters in Eq.~(\ref{BsphiK}) are written with tildes compared to
those in Eq.~(\ref{BdKK}). There are several reasons. First, there is
a different spectator quark in the two decays. Second, the decay $\bs
\to \phi\ks$ receives additional contributions from electroweak
penguins and Zweig-suppressed gluonic penguins. And third, the decay
$\bd \to K^0 \kbar$ has two pseudoscalars in the final state, while
$\bs \to \phi\ks$ has a vector and a pseudoscalar. Because of these
differences, we expect that the parameters in Eq.~(\ref{BdKK}) are not
equal to their counterparts with tildes in Eq.~(\ref{BsphiK}).

There are thus a total of 8 unknowns in the two amplitudes: $\beta$,
$\beta'$, ${\cal P}_{cu}$, $\tilde{\cal P}_{cu}$, ${\cal P}_{tu}$,
$\tilde{\cal P}_{tu}$, $\delta_{cu}-\delta_{tu}$, and
$\tilde\delta_{cu}-\tilde\delta_{tu}$. However, there are (as usual)
only 7 measurements: the magnitudes and relative phase of $A_d^{\sss
  KK}$ and ${\bar A}_d^{\sss KK}$, the magnitudes and relative phase
of $A_s^{\sss \phi K}$ and ${\bar A}_s^{\sss \phi K}$, and the weak
phase in $\bd$-$\bdbar$ mixing, $\beta$.

The number of theoretical unknowns can be made equal to the number of
measurements by making an assumption. In Ref.~\cite{KLY} it is assumed
that $r=\tilde{r}$, where $r \equiv {\cal P}_{cu}/{\cal P}_{tu}$ and
$\tilde{r} \equiv \tilde{\cal P}_{cu}/\tilde{\cal P}_{tu}$. The
uncertainty on this assumption is estimated to be fairly small:
\beq
{r - \tilde{r} \over r} \simeq 20\% ~.
\eeq
Thus, taking $r\simeq \tilde{r}$ is a reasonably good approximation.
With this assumption, we now have 7 measurements in 7 unknowns, and we
can therefore solve for $\beta$ and $\beta'$ separately, up to
discrete ambiguities. (The explicit solution is given in
Ref.~\cite{KLY}.) A comparison of $\beta$ and $\beta'$ may then reveal
the presence of new physics in the $b\to d$ FCNC.

\subsection{Isospin Analysis of $B \to \pi\pi$}

In Sec.~2.2 we discussed how an isospin analysis of $B\to \pi\pi$
decays allows one to remove the penguin pollution and cleanly extract
$\alpha$. However, there are not enough measurements to obtain further
information about the remaining theoretical parameters. We had
\bea
{1\over \sqrt{2}} \, A^{+-} & = & T^{+-} e^{i \delta^{+-}}
e^{-i \alpha} + P e^{i \delta_{\sss P}} e^{-i\thetaNP} ~, \nn\\
{1\over \sqrt{2}} \, {\bar A}^{+-} & = & T^{+-} e^{i \delta^{+-}}
e^{i \alpha} + P e^{i \delta_{\sss P}} e^{i\thetaNP} ~.
\eea
Defining
\beq
2 \alpha_{eff} \equiv {\rm Arg} ({A^{+-}}^* \Abar^{+-}) ~,
\eeq
we therefore have 3 measurements ($|A^{+-}|$, $|{\bar A}^{+-}|$,
$2\alpha_{eff}$) in 4 unknowns ($T^{+-}$, $P$, $\delta^{+-} -
\delta_{\sss P}$, $\thetaNP$). 

Suppose that we knew the value of $P/T^{+-}$. This could come, for
example, from a theoretical estimate or a lattice calculation. In this
case, we can solve for $\thetaNP$. One can derive the following
expressions \cite{Charles}:
\bea
P^2 & = & { |A^{+-}|^2 + |{\bar A}^{+-}|^2 
- 2 |A^{+-}| |{\bar A}^{+-}| \cos(2 \alpha - 2 \alpha_{eff}) 
\over 8 \sin^2 (\alpha-\thetaNP)} ~, \nn\\
\left(T^{+-}\right)^2 & = & { |A^{+-}|^2 + |{\bar A}^{+-}|^2 
- 2 |A^{+-}| |{\bar A}^{+-}| \cos(2 \thetaNP - 2 \alpha_{eff}) 
\over 8 \sin^2 (\alpha-\thetaNP)} ~.
\eea
Thus, if we assume a particular value of the ratio $P/T^{+-}$, these
expressions allow us to derive $\thetaNP$ in terms of known
quantities. And we again stress that the assumption about the value of
$P/T^{+-}$ holds only within a particular parametrization, thus
lifting the CKM ambiguity.

\subsection{Angular Analysis of $\bd(t) \to K^* {\bar K}^*$}

In Sec.~2.4, the decay $\bd \to K^* {\bar K}^*$ was discussed in the
context of an angular analysis. The helicity amplitudes were
\bea 
{\cal A}_\lambda & = & 
{\cal P}_{cu}^\lambda \, e^{i\delta_{cu}^\lambda} \, e^{i \beta} 
+ {\cal P}_{tu}^\lambda \, e^{i\delta_{tu}^\lambda} \, e^{-i\thetaNP}
~, \nn\\
{\bar{\cal A}}_\lambda & = & 
{\cal P}_{cu}^\lambda \, e^{i\delta_{cu}^\lambda} \, e^{-i \beta} 
+ {\cal P}_{tu}^\lambda \, e^{i\delta_{tu}^\lambda} \, e^{i\thetaNP} ~,
\label{AAbaramps}
\eea
where $\lambda = 0, \perp, \|$. The helicity analysis allows the
extraction of the magnitudes and relative phases of all the ${\cal
  A}_\lambda$ and ${\bar{\cal A}}_\lambda$ amplitudes.

In order to extract $\thetaNP$, it is possible to use one of the
techniques described in the previous subsections. However, one might
also consider an alternative method involving the strong phases.
Defining $\delta^\lambda \equiv \delta_{cu}^\lambda -
\delta_{tu}^\lambda$, suppose we assume that the $\delta^\lambda$'s
are the same for all helicities. This assumption is somewhat in the
spirit of Bander, Silverman and Soni (BSS) \cite{BSS}, in which the
strong phases arise principally from the absorptive parts of the loop
diagrams, and so are independent of the helicity of the final state.

We can then solve for $\thetaNP$ as follows. We define the measurable
quantity $\varphi_{eff}^\lambda$ as
\beq
2 \varphi_{eff}^\lambda \equiv {\rm Arg} ({\cal A}_\lambda^* 
{\bar{\cal A}}_\lambda \, \zeta_\lambda^{*2}) ~.
\eeq
Then we can solve for $\delta^\lambda$ as a function of $\thetaNP$
\cite{Charles}:
\bea
\tan\delta^\lambda & = & { \sin(\beta+\thetaNP) 
\left( |{\cal A}_\lambda|^2 - |{\bar{\cal A}}_\lambda|^2 \right) \over
\cos(\beta+\thetaNP) 
\left( |{\cal A}_\lambda|^2 + |{\bar{\cal A}}_\lambda|^2 \right)
- 2 |{\cal A}_\lambda| |{\bar{\cal A}}_\lambda| 
\cos(2 \varphi^\lambda_{eff} + \beta - \thetaNP) } \nn\\
& = & { \tan(\beta+\thetaNP) \Sigma_{\lambda\lambda} \over
\Lambda_{\lambda\lambda} - 
\sqrt{\Lambda_{\lambda\lambda}^2 - \Sigma_{\lambda\lambda}^2} 
\left(\cos(2 \varphi_{eff}^\lambda + 2\beta) + \tan(\beta+\thetaNP) 
\sin(2 \varphi_{eff}^\lambda + 2\beta) \right) } ~,
\label{BSSsolution}
\eea
where $\Lambda_{\lambda\lambda}$ and $\Sigma_{\lambda\lambda}$ are
defined in Eqs.~(\ref{Lambda11}) and (\ref{Sigma11}). Assuming that
$\delta^\lambda = \delta^\sigma$, where $\lambda$ and $\sigma$
represent different helicities, the above equation allows one to solve
for $\thetaNP$ in terms of measurable quantities only. Of course, the
solution will contain discrete ambiguities, but it will still be
possible to establish whether $\thetaNP$ is nonzero.

We must point out here that, although the assumption of a common
$\delta^\lambda$ for all helicity amplitudes does allow us to obtain
$\thetaNP$, its theoretical justification is problematic. From
Eqs.~(\ref{uelim}) and (\ref{uelimhelicity}), recall that
\bea
{\cal P}_{cu}^\lambda \, e^{i\delta_{cu}^\lambda} & = &
(|P_c^\lambda| e^{i\delta_c^\lambda} - |P_u^\lambda| e^{i\delta_u^\lambda}) 
|V_{cb}^* V_{cd}| ~, \nn\\
{\cal P}_{tu}^\lambda \, e^{i\delta_{tu}^\lambda} & = &
(|P_t^\lambda| e^{i\delta_t^\lambda} - |P_u^\lambda| e^{i\delta_u^\lambda}) 
|V_{tb}^* V_{td}| ~. 
\eea
In the BSS calculation \cite{BSS}, the details of the calculations of
the penguin diagrams are independent of the helicity of the final
state. In particular, the strong phases $\delta_i^\lambda$ ($i=u,c,t$)
are in fact $\lambda$-independent. Furthermore, the factors
$|P_i^\lambda|$ ($i=u,c,t$) can each be written as a
$\lambda$-independent penguin piece multiplied by a {\it common}
$\lambda$-dependent matrix element. In this case, it is
straightforward to verify that, indeed, the strong phase
$\delta^\lambda$ is the same for all helicity states. 

However, there is also a problem: the only $\lambda$ dependence of the
amplitudes ${\cal A}_\lambda$ and ${\bar{\cal A}}_\lambda$ in
Eqs.~(\ref{AAbaramps}) is the presence of this overall multiplicative
common matrix element. And this matrix element cancels in all ratios
involving ${\cal A}_\lambda$ and/or ${\bar{\cal A}}_\lambda$. In
particular, ratios such as $\Sigma_{\lambda\lambda} /
\Lambda_{\lambda\lambda}$ are in fact {\it $\lambda$-independent}. But
this implies that the right-hand side of Eq.~(\ref{BSSsolution}) is
actually independent of $\lambda$, so that the technique does not
work.

The only possible loophole in the above argument is that the BSS
calculation makes use of factorization. If non-factorizable effects
are large --- and they may well be for penguin diagrams --- then there
may not be a common $\lambda$-dependent matrix element for each of the
$|P_i^\lambda|$ factors. In this case, the helicity dependence of
${\cal A}_\lambda$ and ${\bar{\cal A}}_\lambda$ is considerably more
complicated than a simple multiplicative factor. On the other hand, in
general the strong phases $\delta^\lambda$ will then not all be equal.

Thus, in order to justify the assumption of a common $\delta^\lambda$
for all helicity amplitudes, one must hope that non-factorizable
effects exist which give different matrix elements for the various
internal-quark contributions, but nevertheless give the same
$\delta^\lambda$ for all helicities. Although this is a logical
possibility, it must be admitted that it seems implausible.

\section{Conclusions}

In the coming years, many measurements will be made of CP violation in
the $B$ system. Hopefully these measurements will reveal the presence
of new physics. Although this in itself will be very exciting, we will
then want to know what kind of new physics it is, and how it has
affected the CP asymmetries.

New physics generally can affect CP-violating asymmetries through its
effects on loop-level processes, such as $B^0$-${\overline{B^0}}$
mixing or penguin decays. It is then useful to categorize these
effects as belonging to the $b\to s$ or the $b\to d$ flavour-changing
neutral current (FCNC). Although there are several ways to cleanly
test for the presence of new physics in the $b\to s$ FCNC, it is not
so easy to do this for the $b\to d$ FCNC.

In the SM, the weak phase in $\bd$-$\bdbar$ mixing is the same as that
found in the $t$-quark contribution to the $b\to d$ penguin amplitude.
In the presence of new physics, these two phases may be different. The
question then is: can one measure the phase of the penguin amplitude?
In this paper we have shown that it is {\it not} possible to cleanly
measure this phase. The reason is essentially the following. There are
three contributions to the $b \to d$ penguin, coming from internal
$u$, $c$ and $t$-quarks. However, due to the unitarity of the CKM
matrix, it is always possible to write any one of these contributions
in terms of the other two. We call this the ``CKM ambiguity.'' It is
therefore not possible to isolate the $t$-quark contribution, and so
one cannot cleanly measure its weak phase.

We have explicitly analyzed several methods which could conceivably
have been used to try to obtain the weak phase of the $t$-quark
contribution to the $b\to d$ penguin, and found that, indeed, there is
not enough information to extract the phase of the $t$-quark penguin.

However, in performing this analysis, we have also obtained an
interesting result: in all cases there is one more theoretical
(hadronic) unknown than there are measurements. Thus, the addition of
a single assumption about the hadronic parameters, which removes the
CKM ambiguity, allows the extraction of the weak phase of the
$t$-quark penguin. This can then be used to test for the presence of
new physics in the $b\to d$ FCNC. We have given several examples of
methods, along with the corresponding assumptions, in which this can
be done.

\bigskip
\medskip
\noindent
Note Added: while we were writing this paper, we received a paper by
R. Fleischer \cite{Fleischer} which discusses some of these same
issues for the specific case of the angular analysis in $B \to VV$
decays.

\section*{\bf Acknowledgments}

We thank Helen Quinn for helpful correspondence. R.S. thanks D.L. for
the hospitality of the Universit\'e de Montr\'eal, where this work was
initiated. The work of D.L. was financially supported by NSERC of
Canada and FCAR du Qu\'ebec.


\begin{thebibliography}{99}
\bibitem{Charles} Expressions very similar to these, with the
  assumption that $\thetaNP=0$, were first derived in J. Charles,
  \prd{59}{99}{054007}.
\end{thebibliography}
\end{document}